\documentclass[11pt,notitlepage]{article}

\usepackage[margin=1in]{geometry}
\usepackage{amsmath} 
\usepackage{amsfonts}
\usepackage{graphicx}
\usepackage{upgreek}
\usepackage{latexsym,slashed}
\usepackage{amssymb}
\usepackage{dcolumn}
\usepackage{cancel}
\usepackage{bm}
\usepackage{mathrsfs}
\usepackage{tikz}
\usetikzlibrary{shapes,arrows}

\tikzstyle{line} = [draw, -latex']
\tikzstyle{cloud} = [draw, ellipse, node distance=2cm, minimum height=1em]

\title{A phase space analysis for nonlinear bulk viscous cosmology.}

\author{G. Acquaviva\footnote{acquavivag@unizulu.ac.za}, A. Beesham\footnote{beeshama@unizulu.ac.za}\\ Department of Mathematical Sciences, University of Zululand\\ Private Bag X1001, Kwa-Dlangezwa 3886, South Africa}
\date{}

\begin{document}

\maketitle

\begin{abstract}
 We consider a Friedmann-Robertson-Walker spacetime filled with both viscous radiation and nonviscous dust.  The former has a bulk viscosity which is proportional to an arbitrary power of the energy density, {\it i.e.} $\zeta \propto \rho_v^{\nu}$, and viscous pressure satisfying a nonlinear evolution equation.  The analysis is carried out in the context of dynamical systems and the properties of solutions corresponding to the fixed points are discussed.  For some ranges of the relevant parameter $\nu$ we find that the trajectories in the phase space evolve from a FRW singularity towards an asymptotic de Sitter attractor, confirming and extending previous analysis in the literature.  For other values of the parameter, instead, the behaviour differs from previous works.
\end{abstract}

%\pacs{98.80.-k, 95.36.+x}

\section*{Introduction}

The $\Lambda$CDM model, which is essentially a homogeneous and isotropic Friedmann--Robertson--Walker (FRW) spacetime filled with both pressureless dust and a cosmological constant, fits very well the evolution of the observable Universe from structure formation to the present accelerated expansion.  The possibility of deviating from such a simple yet effective model has been widely considered throughout the literature, {\it e.g.} by taking into account less symmetric backgrounds or by introducing less ideal matter sources.  For instance, in the context of generalizing the type of fluid that sources the cosmological evolution, one could think of introducing dissipative terms in the energy-momentum tensor, in order to take into account nonequilibrium effects in the fluid(s).  A first approach for describing nonequilibrium thermodynamic effects in a relativistic context was given by Eckart \cite{eckart:1940}.  Unfortunately such a theory presents noncausal features, admitting superluminal propagation of the dissipative signals: this pathology lead other authors to seek for a second-order, causal extension, which could be identified in the so-called Israel--Stewart's theory (IS) \cite{is:1979} (different approaches to the problem have been put forward, {\it e.g.} \cite{ger:1990,cart:1991}, and have been shown to agree with IS not far from equilibrium).  In the present paper we work along these lines, by including in the dynamics of an expanding metric a nonequilibrium contribution to the equilibrium pressure $p$ of the fluid, in the form of a bulk viscous pressure $\Pi$ (some previous studies on the topic can be found {\it e.g.} in \cite{bar:1988,bar:1986,pad:1987,bar:1990,maart:1995,fabris:2006,coli:2007,piat:2011,brev:2011,ave:2013}).  It is well known that, in considering a homogeneous and isotropic background, other dissipative processes with directional character -- heat transfer and shear viscosity -- do not contribute to the dynamics.  The model governing the evolution of bulk viscous pressure that we choose to consider is a nonlinear extension of Israel-Stewart's model (nIS from now on), presented for the first time in \cite{maartens:1997} and further analysed in \cite{chim:1997,ours:2014}.  Apart from the relaxational time $\tau$ typical of IS causal theory, the nIS introduces a characteristic time $\tau_*$ for nonlinear effects such that, when $\tau_*\rightarrow0$, one recovers IS.  The phenomenological nature of such a model has been made clear in the original work and the general conditions underlying the hydrodynamic description (at the foundation of all the previously mentioned models) are also understood: for instance, one has to be aware that the requirement that the mean interaction time $t_c$ between fluid particles be less than the time scale $\theta^{-1}$ set by cosmological expansion (where $\theta$ is the expansion scalar) could cease to hold in the case of accelerated expansion.\\

The main task of the present work, carried out by means of a dynamical system approach, is to describe the space of solutions of an expanding FRW spacetime in presence of a double-fluid source: a pressureless dust component and a viscous radiation component whose bulk pressure is governed by the nIS model.  We do not necessarily focus on the viability of the results in terms of observational consistency; rather we choose to explore in a more general way the impact of the specific nonlinear model on the past and future dynamics.\\

We first present the background setting and the model in Sec.\ref{one}; we then analyse the corresponding dynamical system in Sec.\ref{two}, where critical points of the system are identified and their stability assessed; in Sec.\ref{gen} we give interpretation of the results in terms of cosmological models and finally we give some concluding comment in Sec.\ref{conc}.

\section{The model}\label{one}

Our analysis takes place in standard general relativity, hence Einstein's field equations (EFE)
\begin{equation*}
 G_{\mu \nu} = \kappa\, T_{\mu \nu}
\end{equation*}
are considered a valid starting point.  From now on we will adopt the convention $c=\kappa=1$.  The assumptions of homogeneity and isotropy justify the choice of a Friedmann-Robertson-Walker (FRW) background metric
\begin{equation*}
 ds^2 = -dt^2 + a^2(t) \left( dr^2 + r^2\, d\theta^2 + r^2\, \sin^2\theta\, d\phi^2 \right)
\end{equation*}
and specify in this way the form of the left hand side of EFE.  Through the scale factor $a(t)$ we can define the expansion scalar $\theta=3 \dot{a}/a$, where dot represents the derivative w.r.t. cosmic time $t$.  The right hand side of EFE is built by choosing a noninteracting mixture of dust, with energy density $\rho_d$, and viscous radiation, with energy density $\rho_v$.  Using the equation of state for radiation $p_v=\frac{1}{3}\rho_v$, the latter fluid has an effective pressure $p_{eff} = \frac{1}{3}\rho_v + \Pi$, where the first term is the equilibrium part whereas $\Pi$ is the nonequilibrium contribution, {\it i.e.} the bulk viscous pressure.  By specifying the metric and the matter content of the model we are then able to write down the relevant equations for the system.  First of all, the Friedmann constraint equation coming from EFE,
\begin{equation}
 \rho_d + \rho_v = \frac{1}{3}\theta^2\ ,\label{constr}
\end{equation}
allows us to disregard the evolution of the dust component and hence to focus only on the evolution for the energy density of the viscous fluid: by imposing energy-momentum conservation and making use of the constraint we have that
\begin{equation}
 \dot{\rho}_v = -\theta \left( \frac{4}{3} \rho_v + \Pi \right)\ .\label{conserv}
\end{equation}
Moreover, from EFE we obtain the Raychaudhuri equation
\begin{equation}
 \dot{\theta} = -\frac{1}{2}\theta^2 - \frac{1}{2} \left( \rho_v + 3 \Pi \right)\ ,\label{ray}
\end{equation}
where we made use again of eq.\eqref{constr}.

The introduction of a viscous pressure begs for a specification of its nature and behaviour through some evolution equation.  The model of evolution we will consider here has been proposed in \cite{maartens:1997} and analysed also in \cite{chim:1997} for a single fluid and in \cite{ours:2014}:
\begin{equation}
 \tau  \dot{\Pi} = -\zeta\,  \theta -\Pi \left(1+\Pi\,  \frac{\tau_*}{\zeta }\right)^{-1} -\frac{1}{2}\, \Pi\, \tau \left[\theta +\frac{\dot{\tau}}{\tau}-\frac{\dot{\zeta}}{\zeta}-\frac{\dot{T}}{T}\right]\ .\label{nonlin}
\end{equation}
This is a nonlinear extension of IS equation.  The nonlinear character of the nIS model is expressed by the term in round brackets and it is governed by a relaxational time $\tau_*$: if such characteristic time $\tau_*\rightarrow0$, we recover IS theory.  The nonlinear term itself is inversely proportional to the entropy production rate in the fluid and hence, for thermodynamic consistency, has to be positive.  The elements that appear in eq.\eqref{nonlin} are defined as follows:
\begin{itemize}
 \item bulk viscosity: $\zeta = \zeta_0 \rho_v^{\nu}$, with $\zeta_0>0$ and $\nu\geq0$,
 \item linear relaxational time: $\tau=\zeta/(\gamma\, c_b^2\, \rho_v)$,
 \item nonlinear relaxational time: $\tau_* = k^2 \tau$, with real $k$,
 \item temperature: $T= T_0 \rho_v^{(\gamma-1)/\gamma}$, by assumption of barotropicity of the fluid.
\end{itemize}
We recall that $c_b$ is the dissipative part of the speed of sound, to be taken into account along with the adiabatic part $c_s$ in the total expression for the speed of sound $V^2=c_s^2+c_b^2$.  From the definition of adiabatic speed of sound and the condition $V^2\leq1$, it is easy to find the bound $c_b^2 \leq 2/3$.\\

\section{The Dynamical System}\label{two}

In order to write the evolution equations in the form of an autonomous system, we define the following dimensionless variables:
\begin{equation}\label{def}
\Omega=\frac{3 \rho_v}{\theta^2}\ \ \ ,\ \ \ \tilde{\Pi}=\frac{3 \Pi}{\theta^2}\ \ \ \text{and}\ \ \ \text{Z}=\frac{\zeta_0}{3^{\nu}\theta^{1-2\nu}}.
\end{equation}
We define also a new time variable $\tau$ through $dt/d\tau=3/\theta$, whose action on the variables will be denoted by a prime.  Evolution towards increasing $\tau$ is then related to the evolution of an expanding universe.  By deriving the definitions in eqs.\eqref{def} w.r.t. $\tau$ and making use of eqs.\eqref{constr}--\eqref{nonlin}, we can write down the autonomous system:
\begin{align}
 \Omega' &= -(1- \Omega)\left[ \Omega + 3\, \tilde{\Pi} \right]\label{auto1}\\
 \tilde{\Pi}' &= - 4\, c_b^2\, \Omega \left[ 1+\frac{\tilde{\Pi}}{3\ \text{Z}\, \Omega^{\nu-1}} \left( \Omega+\frac{3\, k^2}{4\, c_b^2}\ \tilde{\Pi} \right)^{-1} \right]\nonumber\\
  &\ \ \ + 3 \frac{\tilde{\Pi}^2}{\Omega}\left(\Omega - \frac{5}{8}\right) - (1-\Omega)\tilde{\Pi}\label{auto2}\\
 \text{Z}' &= - \text{Z} \left( \nu - \frac{1}{2} \right)\left[ \Omega + 3\, ( \tilde{\Pi} + 1 ) \right]\ ,\label{auto3}
\end{align}
Due to the constraint eq.\eqref{constr} we have $\Omega\in (0,1]$ and $\tilde{\Pi}\in \left( -\Omega\, 3c_b^2/4k^2 , +\infty \right)$ from the requirement of positive entropy production.  It is immediately clear from eq.\eqref{auto2} that the system is ill-defined in the subspaces Z$\ =0$ and $\Omega=0$ so, apart from locating the critical points, we will need to study the behaviour of the system in a neighborhood of such planes in the phase space: this will be done through a numerical analysis.  Note also that the singular behaviour in Z $=0$ will prevent this variable to change sign during the evolution: in the following we will focus on the part where Z $>0$.

In the case when $\nu=1/2$, eq.\eqref{auto3} gives Z $=constant$, as can be seen also from the definition in eq.\eqref{def}: the dimensionality of the system is thus reduced from 3 to 2 and the analysis is formally equivalent to the one carried out in \cite{ours:2014}.  With regard to this case we will just report some general considerations in Sec. \ref{special}.

\subsection{Finite analysis ($\nu\neq1/2$)}\label{fin}

We start by locating and analyzing the critical points with finite values of the coordinates in the phase space.\
\\

As long as $\nu\neq1/2$, we note that a generic finite critical point $\{\Omega_c,\tilde{\Pi}_c,\text{Z}_c \}$ with Z$_c\neq0$, if it exists, describes a de Sitter model because, from the definition of Z,
\begin{equation}
 \theta = \left( \frac{3^{\nu}}{\text{Z}_c\, \zeta_0} \right)^{\frac{1}{2\nu-1}}\equiv\theta_0\label{dstheta}
\end{equation}
and constant.  As a consequence the scale factor of the model in that point will be $\propto \text{Exp}\left( \theta_0\, t/3 \right)$.  The energy density and viscous pressure in such a model would attain, respectively, the constant values
\begin{equation}\label{finiteval}
\frac{1}{3}\, \Omega_c\, \theta_0^2\ \ \ \ \text{and}\ \ \ \ \ \frac{1}{3}\, \tilde{\Pi}_c\, \theta_0^2\ .
\end{equation}\\

In the invariant set $\Omega=1$ ({\it i.e.} where $\Omega'=0$) we find a finite critical point of this kind:
\begin{equation}
 P_0 = \left\{\, 1\, ,\, -\frac{4}{3}\, ,\, \frac{4\, c_b^2}{9 \left( c_b^2 - 1/2 \right)\, \left( 1-k^2/c_b^2 \right)}\,  \right\}
\end{equation}
Requiring positivity of Z imposes either $\{c_b^2>1/2\ \wedge\ k^2<c_b^2\}$ or $\{c_b^2<1/2\ \wedge\  k^2>c_b^2\}$, but only in the former case $P_0$ lies in the physical region of the phase space given by the constraints on the variables.  Thus, considering the bound on $c_b$ found previously, the existence of this critical point is subject to the condition
\begin{equation}
1/2 < c_b^2 \leq 2/3\ .\label{vbound}
\end{equation}
Also in \cite{maartens:1997} possibility of exponential inflation was found in this range of velocities.  The location of $P_0$ does not depend on $\nu$, but the stability properties depend crucially on the exponent.  By studying the linearized system around the point we conclude that $P_0$ is a stable attractor for $\nu<1/2$ and an unstable saddle for $\nu>1/2$.  In the following we will assume validity of the bound eq.(\ref{vbound}), because the dynamics results to be richer.  We will make some comments on the case $0<c_b^2\leq1/2$ in Sec. \ref{gen}.\\

A second finite critical point can be found by noting that one can solve the equilibrium condition for the system by approaching $\tilde{\Pi}=\text{Z}=0$ in the invariant set $\Omega=1$ along the line $\tilde{\Pi}=-3\, \text{Z}$.  In such a way $P_1 = \{ 1,0,0 \}$ results to be a critical point of the system.  The point $P_1$ is a saddle, so it will act only as a transient stage for some trajectories in the phase space.\\

It is also possible to discover another critical point in the finite phase space if we approach the line $\tilde{\Pi}=\text{Z}=0$ along a generic curve $\tilde{\Pi}\propto\text{Z}^n$ (with $n\geq1$) and then we let $\Omega\rightarrow0$.  From numerical analysis, the point $P_2 = \{ 0,0,0 \}$ found acts either as a saddle (if $0<\nu<1/2$) or as a sink (if $\nu>1/2$).\\
\begin{figure*}
 \begin{center}
 \includegraphics[width=10.5cm]{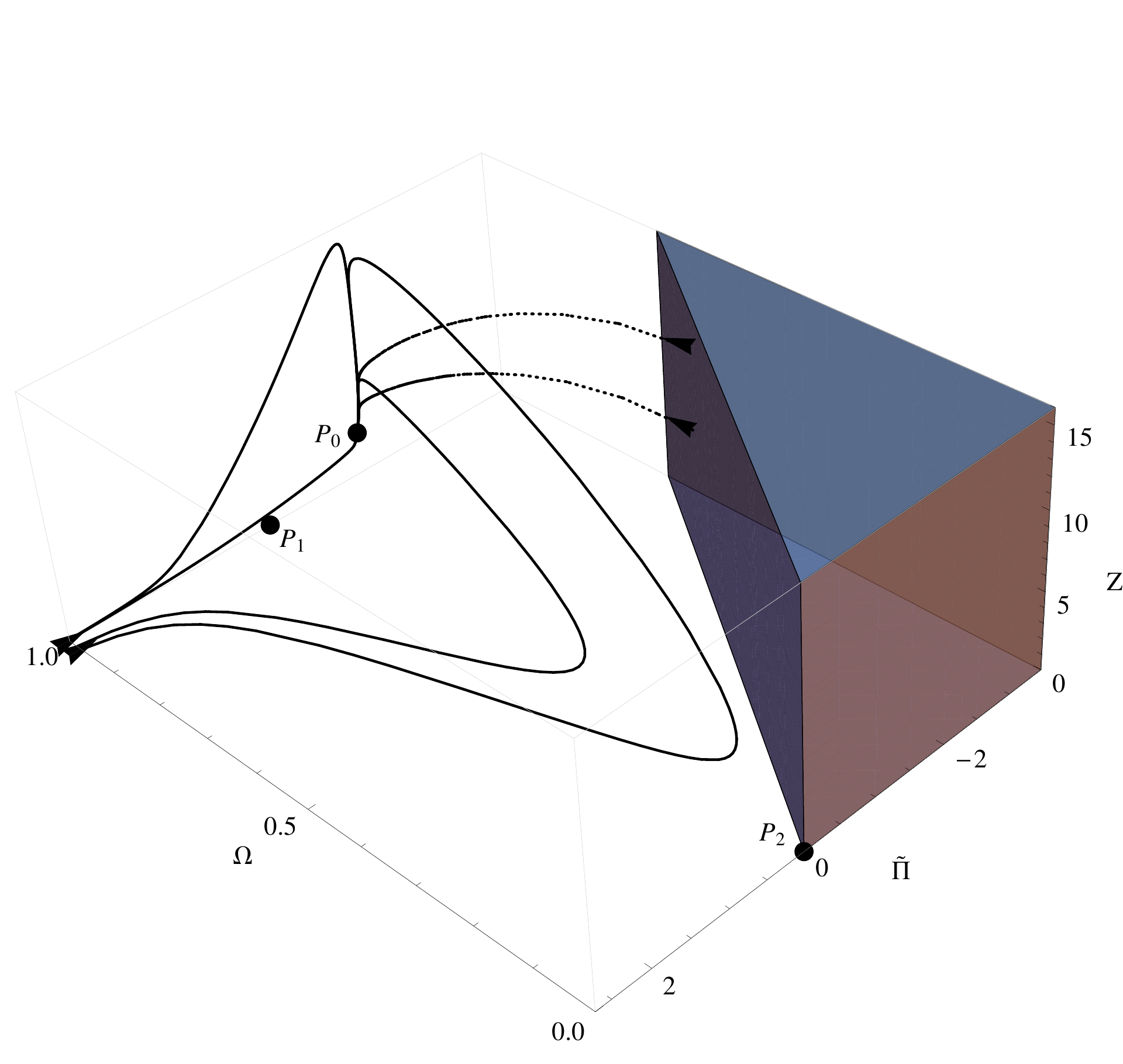}
 \includegraphics[width=10.5cm]{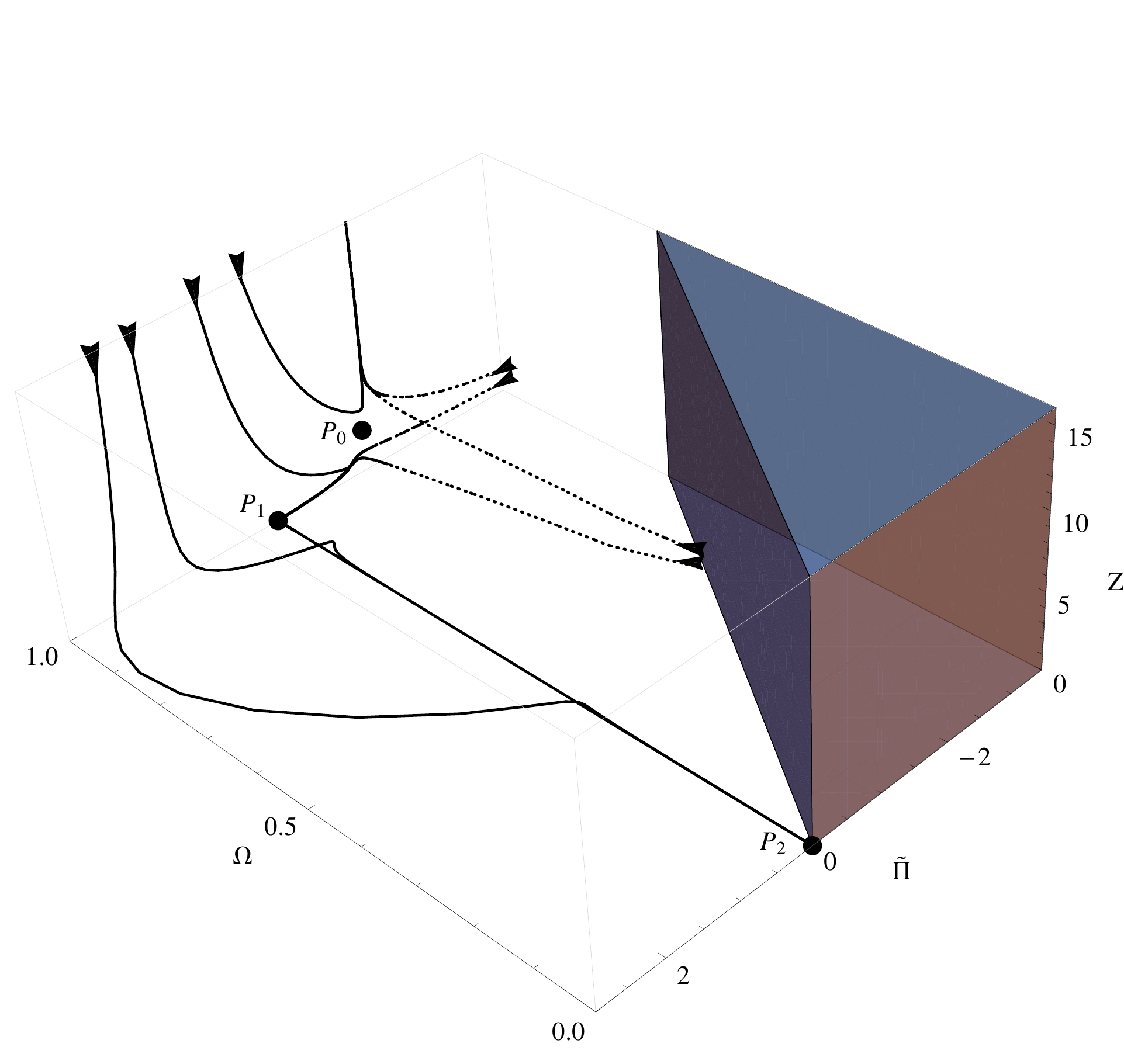}
 \caption{\label{finiteplot} Representative trajectories in the phase space for $c_b=0.8$ and $k=0.4$.  Arrowheads specify the direction of the trajectories.  The location of $P_0$ is specified by a black dot: for $\nu=0$ (left) it is a sink, for $\nu=3/2$ (right) it is a saddle.  The region of negative entropy production is shaded and trajectories emerging from it are dotted.}
\end{center}
\end{figure*}

The stability properties of the finite critical points are summarized in Table \ref{fincritical}.  In Fig.\ref{finiteplot} some representative trajectories in the phase space are plotted, highlighting in this way the impact of the choice of $\nu$ on the overall behaviour.  The shaded part of the phase space is the negative entropy production region and trajectories emerging from it are denoted by dotted lines.  The direction of the trajectories is specified by arrowheads and the locations of the finite critical points are identified by black dots.  From the plots (and from the absence of sources for $\nu\neq1/2$ in Table \ref{fincritical}) it is clear that the asymptotic points play an important role, as we will see in the next section.

\setlength{\extrarowheight}{10pt}
\begin{table}
 \begin{center}
  \begin{tabular}{c||ccccc}
    Point: & $P_0$ & $P_1$ & $P_2$ & $P^+$ & $P^-$\\\hline
    $0\leq\nu<1/2$ & sink & saddle & saddle & (n/a) & (n/a)\\
    $\nu=1/2$ & (n/a) & (n/a) & saddle & source & sink\\
    $\nu>1/2$ & saddle & saddle & sink & (n/a) & (n/a)\\
  \end{tabular}
 \end{center}
\caption{\label{fincritical} Stability of the finite critical points of the system.}
\end{table}

\subsection{Asymptotic analysis ($\nu\neq1/2$)}\label{asy}

The study of the critical points at infinity is carried out by compactifying the phase space.  This is done by redefining the unbounded variables through the following transformation:
\begin{equation}\label{poinc}
 \tilde{\Pi} = \frac{r}{1-r}\, \cos \phi\ \ \ ,\ \ \ \text{Z} = \frac{r}{1-r}\, \sin \phi\ ,
\end{equation}
where $r\in [0,1)$ and $\phi \in [0,\pi]$.  In this way, the directions of divergence of (either one or both) the variables are mapped on the boundary $r=1$.  Note that, because of the lower bounds on $\tilde{\Pi}$ and Z, at infinity the variable $\phi$ is restricted to $[0,\pi/2]$.  By using eq.\eqref{poinc} we obtain a system in the implicit form
\begin{align*}
 \Omega' &= f\left( \Omega, r, \phi \right)\\
 r' &= g\left( \Omega, r, \phi \right)\\
 \phi' &= h\left( \Omega, r, \phi \right)\ .
\end{align*}
The location of the critical points at infinity can be obtained by evaluating first the limit $r\rightarrow 1^-$ and then by finding the values $\{ \Omega_c, \phi_c \}$ that satisfy $\Omega'=\phi'=0$.  The stability of the critical points is then checked through the linearization matrix of the 2-dimensional phase space $\{ \Omega,\phi \}$ at infinity together with the sign of $r'$, which reflects the stability in the radial direction.  We made use of numerical studies in order to assess the stability of some of the asymptotic critical elements.\\  

For $r\rightarrow 1^-$, the system at infinity is given by
\begin{align}
 \Omega' &= -3 \cos\phi\, \frac{(1-\Omega )}{1-r}\label{inf1}\\
 r' &= -\frac{3 \cos\phi}{8 \Omega }\, \cdot\nonumber\\ 
&\cdot\left(5 \cos^2\phi+4 \left(\nu -\frac{3}{2}-2\left(\nu +\frac{1}{2}\right) \cos(2 \phi) \right) \Omega \right)\label{inf2}\\
 \phi' &= \frac{3 \cos\phi}{8 \Omega\, (1-r) }\cos\phi\, \sin\phi \left(5-8\left(\nu +\frac{1}{2}\right) \Omega \right)\ .\label{inf3}
\end{align}

First of all, in the particular case when $\nu=1/8$ we note that the condition $\Omega'=\phi'=0$ identifies the value $\Omega=1, \forall\phi$: thus for this choice of the exponent we find an asymptotic critical line $L_1^{\infty}=\{ 1, \phi \}$ with saddle behaviour.

For $\nu\neq1/8$ there is no critical line $L_1^{\infty}$ but a critical point in $P_1^{\infty}=\{ 1, 0 \}$: from numerical studies this point turns out to be a source for $\nu<1/2$ and a saddle for $\nu>1/2$.\\

Finally, the common factor $\cos\phi$ in eqs.\eqref{inf1} and \eqref{inf3} tells us that the system has a critical line $L_2^{\infty}=\{ \Omega, \pi/2 \}$, which is a saddle for $\Omega\in(0,1)$.  In this line we can identify two particular critical points: $l_2^{\infty}=\{ 1,\pi/2 \}$ and $m_2^{\infty}=\{ 0,\pi/2 \}$.  

The point $l_2^{\infty}$ is equivalent to Z $\rightarrow\infty$ in the invariant set $\Omega=1$.  By solving $\tilde{\Pi}'=0$ for Z with $\Omega=1$ in the original system, it is possible to see that this asymptotic critical point actually corresponds to two vertical asymptotes $\tilde{\Pi}=\pm \frac{4}{3}\sqrt{2}\, |c_b|$ in the invariant set, with different stability features: along the positive asymptote, $l_2^{\infty}$ is a saddle for $\nu<1/2$ and a source for $\nu>1/2$; along the negative asymptote, it is a saddle for $\nu<1/2$ and a sink for $\nu>1/2$.  The degeneracy in this asymptotic point is made apparent in the bottom panel of Fig.\ref{asympplot2}, representing the compactified invariant set: the point's double nature of source and sink is shown for a representative trajectory in a case where $\nu=3/2$.  

The point $m_2^{\infty}$ corresponds to Z $\rightarrow\infty$ in the plane $\Omega=0$.  This point is a saddle $\forall\nu$.

\setlength{\extrarowheight}{10pt}
\begin{table}
 \begin{center}
  \begin{tabular}{c||cccc}
    Point: & $P_1^{\infty}$ & $L_1^{\infty}$ & $l_2^{\infty}$ & $m_2^{\infty}$\\\hline
    $0\leq\nu<1/8$ & source & (n/a) & saddle & saddle\\
    $\nu=1/8$ & source & saddle & saddle & saddle\\
    $1/8<\nu<1/2$ & source$$ & (n/a) & saddle & saddle\\
    $\nu=1/2$ & (n/a) & (n/a) & (n/a) & (n/a)\\
    $\nu>1/2$ & saddle & (n/a) & source/sink & saddle\\
  \end{tabular}
 \end{center}
\caption{\label{asympcritical} Stability of the asymptotic critical elements of the system.}
\end{table}

\subsection{The case $\nu=1/2$}\label{special}

If the exponent in the definition of the bulk viscosity takes the value $1/2$, we have a drastic modification of the system: in fact, in this case the variable Z becomes a constant and the phase space is subject to a reduction in dimensionality.  The dynamics now takes place in 2-dimensional invariant subsets ($\text{Z}'=0$) of the original 3-dimensional problem, one subset for each value of Z.

Defining the constant $\text{Z}_0=\zeta_0/3^{1/2}$, we have the following system of equations:
\begin{align}
 \Omega' &= -(1- \Omega)\left[ \Omega + 3\, \tilde{\Pi} \right]\label{red1}\\
 \tilde{\Pi}' &= - 4\, c_b^2\, \Omega \left[ 1+\frac{\tilde{\Pi}\, \Omega^{1/2}}{3\ \text{Z}_0} \left( \Omega+\frac{3\, k^2}{4\, c_b^2}\ \tilde{\Pi} \right)^{-1} \right]\nonumber\\
  &\ \ \ + 3 \frac{\tilde{\Pi}^2}{\Omega}\left(\Omega - \frac{5}{8}\right) - (1-\Omega)\tilde{\Pi}\label{red2}
\end{align}
The situation is analogous to the case treated in \cite{ours:2014}, the only difference being the initial choice for the bulk viscosity $\zeta$ given in Sec. \ref{one}.  Nevertheless the system presents three critical points, corresponding to three roots of a cubic equation:
\begin{itemize}
 \item $P^+$ is a source
 \item $P^-$ is a sink
 \item $P_2$ is a saddle
\end{itemize}
Both the source and the sink represent polynomially expanding solutions with viscous fluid predominance.  A future de Sitter attractor is possible but a fine tuning of the parameters is needed.  The details in \cite{ours:2014} are qualitatively equivalent to this case and we will not discuss them further.\\

\section{Cosmological evolutions}\label{gen}

In translating the properties of the phase space elements in terms of cosmological models we choose to disregard the trajectories emerging from the boundary of the unphysical region (shaded in gray in all the Figs.).  The main feature of their past stage is obviously the fact that the entropy production rate diverges and the system is not well defined.  Focusing on the rest of the phase space, we analyse the cosmological models corresponding to different ranges of $\nu$.\\

\begin{figure}
 \begin{center}
  \includegraphics[width=10cm]{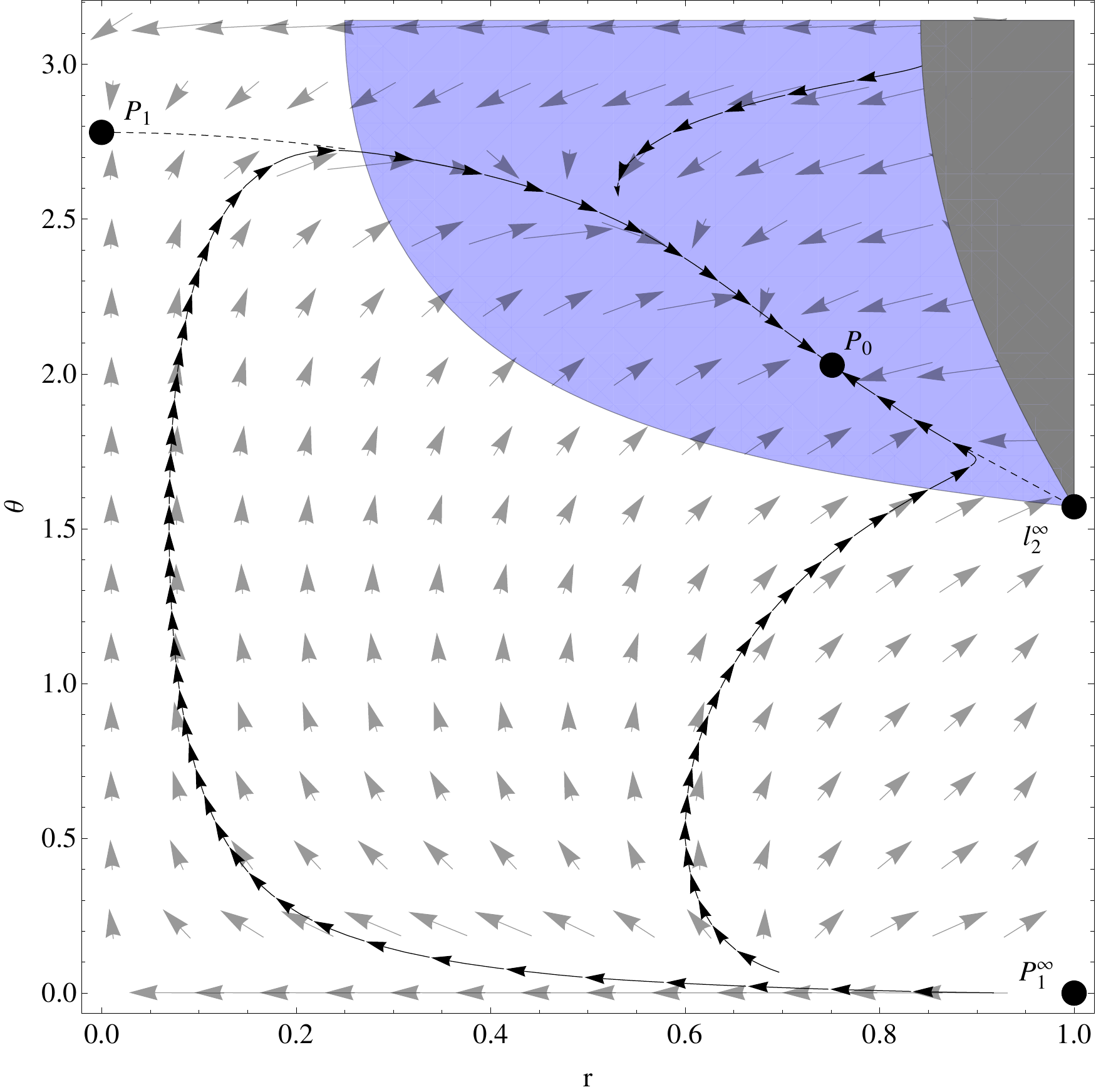}\vspace{0.3cm}
 \includegraphics[width=10cm]{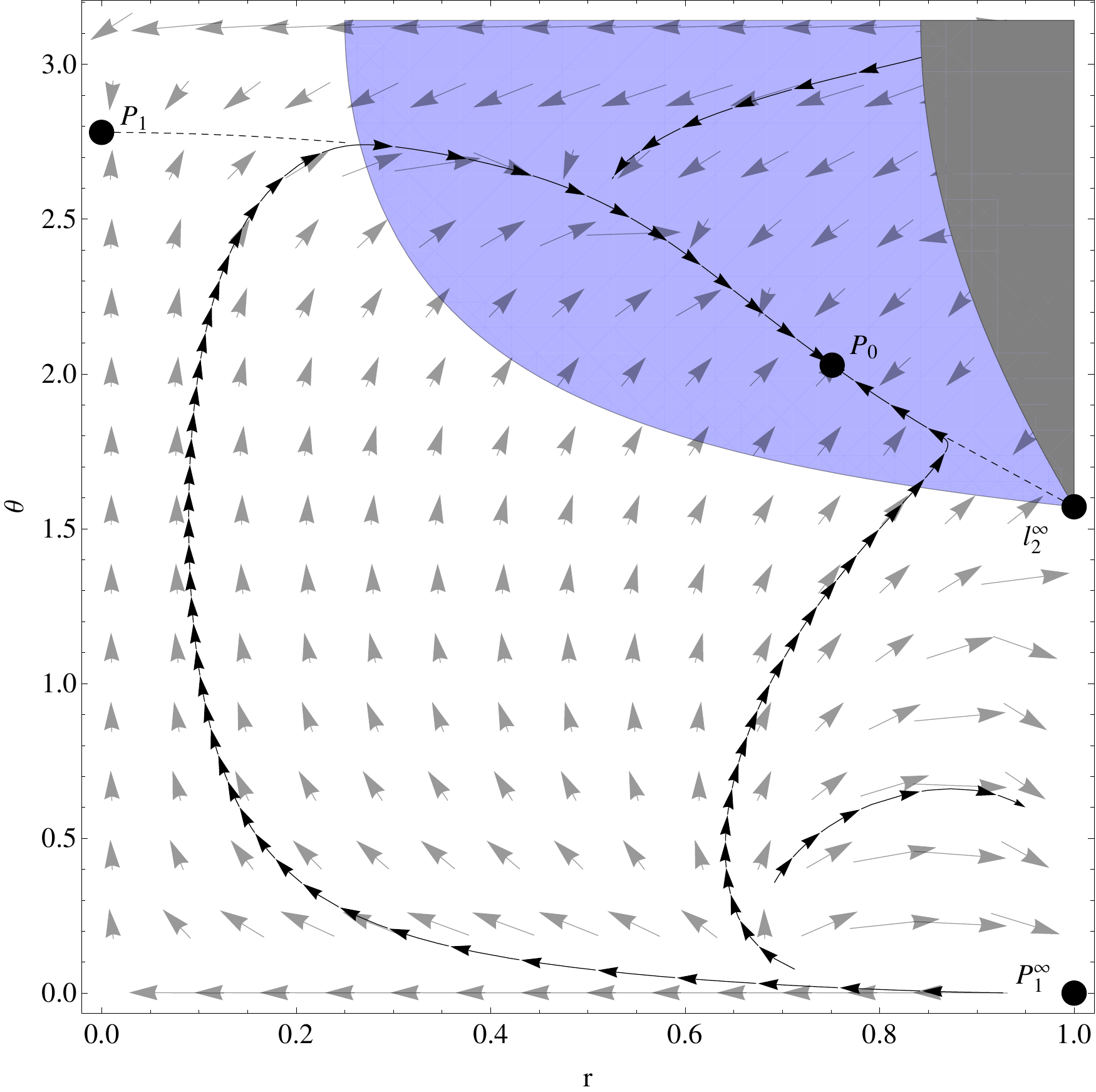}
 \caption{\label{asympplot} Compactified phase space in the invariant set $\Omega=1$ with $c_b=0.8$, $k=0.4$ and $\nu=0$ (top) or $\nu=9/40$ (bottom).  The critical points are denoted by black dots.  Negative entropy production subspace is shaded in gray, while the region of stability in the $\Omega$ direction is shaded in blue.}
\end{center}
\end{figure}

When $0\leq\nu<1/2$, the main elements are a source in $P_1^{\infty}$ and a sink in $P_0$.  Thus, for $a\rightarrow0$ we have a radiation-dominated FRW model, while for $a\rightarrow\infty$ the system evolves towards a stable de Sitter model.  Possible intermediate stages are a nonviscous dust dominated era in $P_2$ and a viscous radiation dominated era in $l_2^{\infty}$, both characterized by polynomial expansion.  Such transient stages are identifiable through the behaviour of the solid trajectories in the left panel of Fig.\ref{finiteplot}.  The compactified invariant set $\Omega=1$ is represented in the top panel of Fig. \ref{asympplot}.  When crossing the value $\nu=1/8$, there is a change in the relative attraction between $P_1^{\infty}$ and $l_2^{\infty}$: if $\nu<1/8$ the direction of the trajectories at infinity is $P_1^{\infty}\rightarrow l_2^{\infty}$, while for $\nu>1/8$ is the opposite.  In the latter case, as a consequence of this change, we also note that the point $P_1^{\infty}$ (apart from being a source) can act as a sink for trajectories lying in the invariant set, as shown in the bottom panel of Fig. \ref{asympplot}.  This means that there exist solutions that are FRW both for  $a\rightarrow0$ and  $a\rightarrow\infty$, but this is obviously an unstable particular situation if the fluid is not purely viscous, because of the instability in the $\Omega$ direction.\\

\begin{figure}
 \begin{center}
\includegraphics[width=10cm]{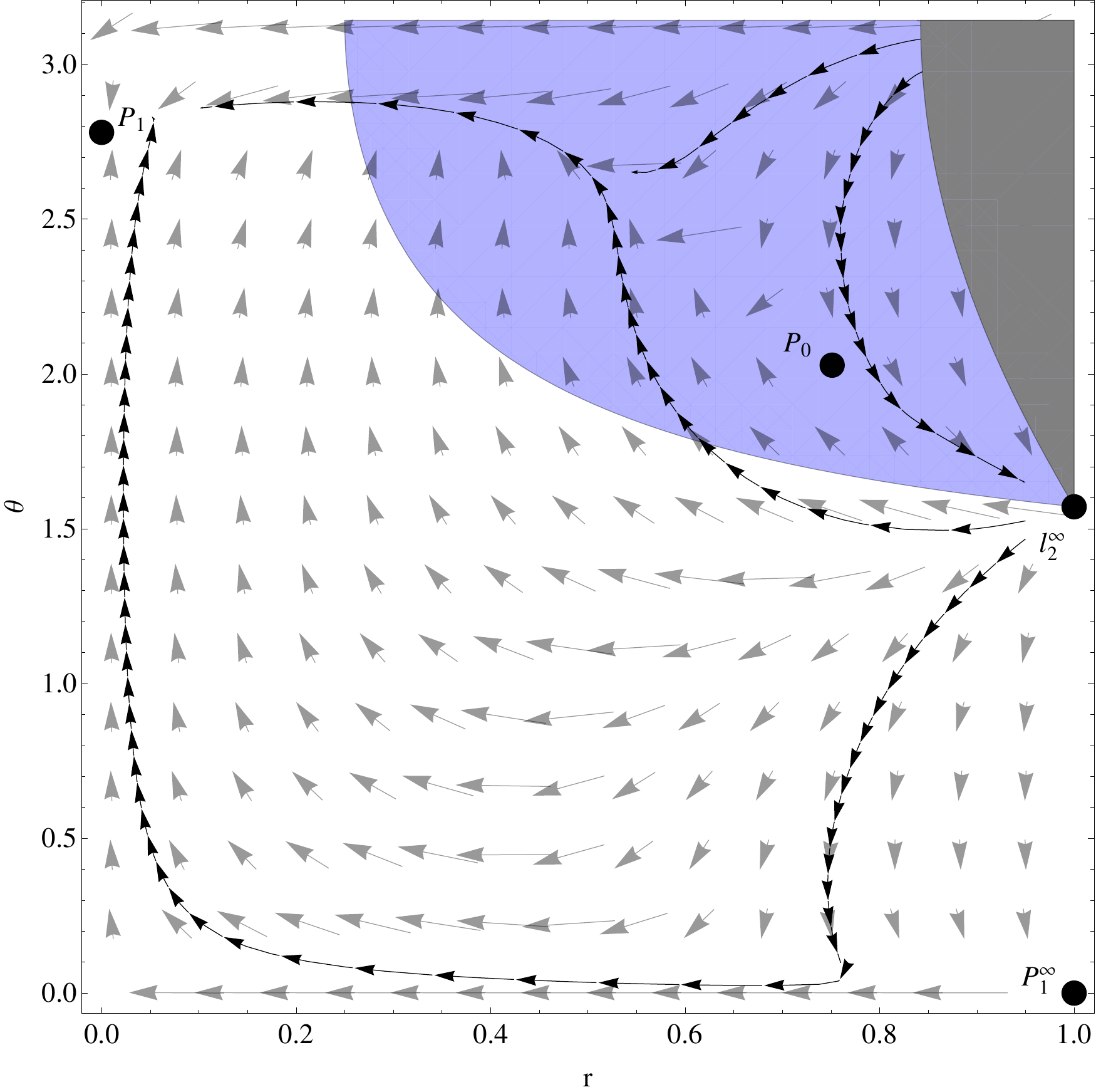}\vspace{0.3cm}
\includegraphics[width=10.16cm]{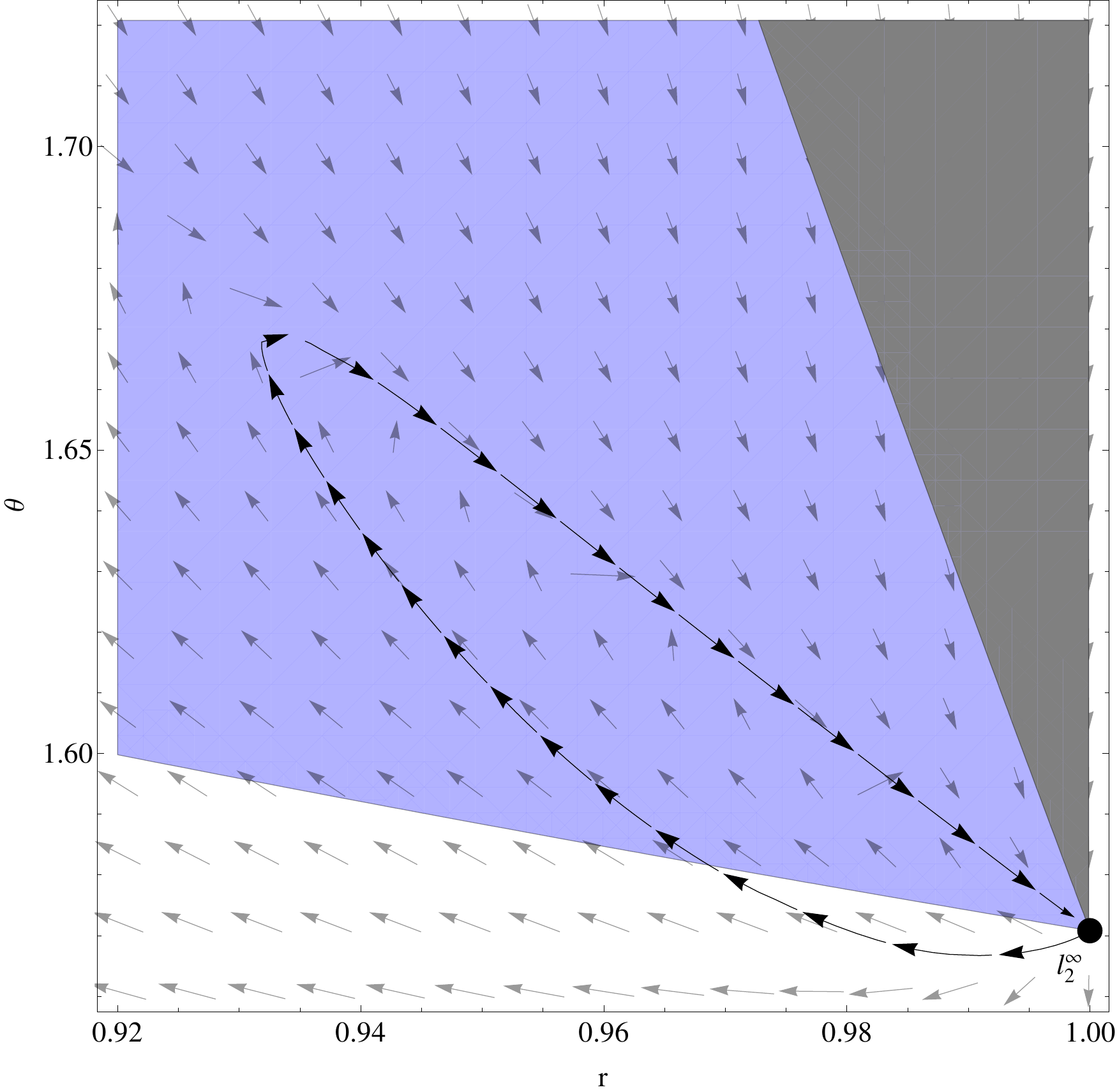}
 \caption{\label{asympplot2} Compactified phase space in the invariant set $\Omega=1$ with $c_b=0.8$, $k=0.4$ and $\nu=3/2$ (top).  The critical points are denoted by black dots.  Negative entropy production subspace is shaded in gray, while the region of stability in the $\Omega$ direction is shaded in blue.  Detail of the region around $l_2^{\infty}$ for the same values of the parameters (bottom): the representative trajectory shows the double nature (source/sink) of the critical point.}
\end{center}
\end{figure}

If $\nu>1/2$ the stability of the critical points changes drastically: the past attractor for the trajectories is now $l_2^{\infty}$ along the positive asymptote, with a scale factor proportional to $(t-t_0)^{\frac{1}{2(1+\sqrt{2}\, |c_b|)}}$ which denotes a slower approach to $a\rightarrow0$ with respect to the pure radiation case.  The main sink for $a\rightarrow\infty$ is $P_2$, which is a polynomially expanding model dominated by non-viscous dust with $a\propto(t-t_0)^{2/3}$.  As we have seen, the point $l_2^{\infty}$ itself is also a sink for some trajectories along the negative asymptote.  The sink in $l_2^{\infty}$ is characterized by $a\rightarrow\infty$, $\rho_v\rightarrow\infty$ and $|\Pi|\rightarrow\infty$, which could represent a future Type I singularity (or Big Rip \cite{noji:2005}) with scale factor $\propto (t_s-t)^{-\frac{1}{2(\sqrt{2}\, |c_b|-1)}}$.  We can calculate the deceleration $q$ and effective EoS $\gamma_{eff}$ parameters in this point, finding

\begin{align}\label{acc}
 q &= 1-2\sqrt{2}|c_b|\\
 \gamma_{eff} &= \frac{4}{3} \left( 1-\sqrt{2}|c_b| \right)\, \label{eff}.
\end{align}

Given the bound on $c_b$ in eq.(\ref{vbound}) we have that $q<-1$ and $\gamma_{eff}<0$, showing an effective phantom behaviour.
As regards the de Sitter model $P_0$, being a saddle in this range of $\nu$, it acts as a possible transient stage.  The situation in the compactified invariant set $\Omega=1$ is represented in the top panel of Fig. \ref{asympplot2} for a value $\nu=3/2$.\\

Some considerations on the bound eq.(\ref{vbound}) are in order.  When $c_b^2\rightarrow1/2$, the de Sitter critical point is pushed towards infinity along the negative asymptote we discussed earlier, coinciding with $l_2^{\infty}$ in the limit.  In the range $0<c_b^2\leq1/2$, the de Sitter critical point $P_0$ is effectively no longer part of our phase space and the point $l_2^{\infty}$ along the negative asymptote inherits its stability properties: it is a sink for $0\leq\nu<1/2$ and a saddle for $\nu>1/2$.  However, the different bound on $c_b$ changes also the cosmological nature of the point, as can be seen from eqs.(\ref{acc})(\ref{eff}): we don't find a phantom model because now $-1<q<1$ and $4/3<\gamma_{eff}<0$.\\

\section{Conclusions}\label{conc}

In the present paper we analysed the phase space of a FRW spacetime filled with both radiation and dust.  The effective pressure of the radiative component splits in an equilibrium and a nonequilibrium part: the latter is a viscous pressure term which satisfies a nonlinear evolution equation, given by eq.(\ref{nonlin}).  This study is a generalization of a previous work \cite{ours:2014} to the case of a general power-law dependence of the bulk viscosity on the energy density, {\it i.e.} $\zeta = \zeta_0\, \rho_v^{\nu}$ with $\nu\geq0$.  We focused mainly on the case where the bound on velocities eq.(\ref{vbound}) is satisfied because in such a range a de Sitter critical point, interesting from the cosmological point of view, is present in the phase space.\\

The system displays interesting features regarding the stability of the critical points (summarized in Tables \ref{fincritical} and \ref{asympcritical}) and hence the evolution of the trajectories in the phase space.  As already known in the literature, the stability properties of the system change when crossing the value $\nu=1/2$.  Our findings for $0\leq\nu<1/2$ are in line with previous studies, such as \cite{bar:1986,bar:1988,fabris:2006,coli:2007}: the evolution for an expanding universe starts with a FRW phase for $a\rightarrow0$ and is asymptotically de Sitter for $a\rightarrow\infty$, a behaviour that survives in this model since the Eckart approach; in our double-fluid system also an intermediate matter dominated stage is possible.  The overall evolution can be schematically represented in the following diagram: \\

\begin{center}
\begin{tikzpicture}[node distance = 1cm, auto]
    % Place nodes
    \node [cloud] (past1) {$P_1^{\infty}$};
    \node [cloud, right of=past1, above of=past1] (one1) {$l_2^{\infty}\, (-)$};
    \node [cloud, right of=past1, below of=past1] (two1) {$P_2$};
    \node [cloud, below of=one1, right of=one1] (fut1) {$P_0$};
    % Draw edges
    \path [line] (past1) -- (one1);
    \path [line] (past1) -- (two1);
    \path [line] (two1) -- (one1);
    \path [line] (one1) -- (fut1);
    \path [line] (two1) -- (fut1);
\end{tikzpicture}
\end{center}
The arrows point in direction of increasing scale factor and the sign in round brackets specifies the asymptote (see Sec. \ref{asy}).  If $\nu>1/2$, differently from other works, we find a viscous radiation-dominated FRW behaviour for $a\rightarrow0$; only in some intermediate stage can we find a de Sitter phase; finally, there are two possible futures for $a\rightarrow\infty$, either a dust-dominated FRW model or a viscosity-driven phantom model:\\

\begin{center}
\begin{tikzpicture}[node distance = 2cm, auto]
    % Place nodes
    \node [cloud] (past2) {$l_2^{\infty}\, (+)$};
    \node [cloud, right of=past2] (one2) {$P_0$};
    \node [cloud, below of=past2, right of=past2] (two2) {$P_1$};
    \node [cloud, below of=one2, right of=one2] (futb) {$P_2$};
    \node [cloud, above of=one2, right of=one2] (futa) {$l_2^{\infty}\, (-)$};
    % Draw edges
    \path [line] (past2) -- (one2);
    \path [line] (past2) -- (two2);
    \path [line] (one2) -- (two2);
    \path [line] (two2) -- (futb);
    \path [line] (one2) -- (futa);
\end{tikzpicture}
\end{center}

\section*{Acknowledgments}
 GA is thankful to the ACRU group at UKZN for the warm hospitality and the useful discussions, and to John D. Barrow for useful comments on the work.  GA is supported by the Postdoctoral Fellowship from the University of Zululand.


\begin{thebibliography}{99}

\bibitem{eckart:1940} C. Eckart, Phys. Rev. {\bf 58}, 919 (1940).

\bibitem{is:1979} W. Israel and J. M. Stewart, Ann. Phys. (N.Y.) {\bf118}, 341 (1979).

\bibitem{ger:1990} R. Geroch and L. Lindblom Phys. Rev. D {\bf41}, 1855 (1990).

\bibitem{cart:1991} B. Cater, Proc. R. Soc. A {\bf433}, 45 (1991).

\bibitem{bar:1986} J. D. Barrow, Phys. Lett. B {\bf180}, 335 (1986).

\bibitem{bar:1988} J. D. Barrow, Nucl. Phys. {\bf310}, 743 (1988).

\bibitem{pad:1987} T. Padmanabhan and S. M. Chitre, Phys. Lett. A {\bf120}, 433 (1987).

\bibitem{bar:1990} J. D. Barrow, {\it The Formation and Evolution of Cosmic Strings}, edited by G.W. Gibbons, S.W. Hawking, and T. Vachaspati (Cambridge University Press, Cambridge, England, 1990), p. 449.

\bibitem{maart:1995} R. Maartens, Classical Quantum Gravity {\bf 12}, 1455 (1995).

\bibitem{fabris:2006} J. C. Fabris, S. V. B. Gon\c{c}alves, and R. de S\'{a} Ribeiro, Gen. Relativ. Gravit., {\bf38} 495 (2006).

\bibitem{coli:2007} R. Colistete, J. Fabris, J. Tossa, and W. Zimdahl, Phys. Rev. D {\bf76}, 103516 (2007).

\bibitem{piat:2011} O. F. Piattella, J. C. Fabris, and W. Zimdahl, J. Cosmol. Astropart. Phys. {\bf11} (2011) 029.

\bibitem{brev:2011} I. Brevik, E. Elizalde, S. Nojiri, and S. D. Odintsov, Phys. Rev. D {\bf 84}, 103508 (2011).

\bibitem{ave:2013} A. Avelino, R. Garcia-Salcedo, T. Gonzalez, U. Nucamendi, and I. Quiros, J. Cosmol. Astropart. Phys. {\bf13} (2013) 012.

\bibitem{maartens:1997} R.  Maartens and V. M\'endez, Phys. Rev. D {\bf55}, 4 (1997).

\bibitem{chim:1997} L. P. Chimento, A. S. Jakubi, V. M\'endez, and R. Maartens, Classical Quantum Gravity {\bf14}, 3363 (1997).

\bibitem{ours:2014} G. Acquaviva and A. Beesham, Phys. Rev. D {\bf90}, 023503 (2014).

\bibitem{noji:2005} S. Nojiri, S. D. Odintsov, and S. Tsujikawa, Phys. Rev. D {\bf71}, 063004 (2005).


\end{thebibliography}
\end{document}